\documentclass[12pt]{article}
\usepackage{amsfonts,amssymb,amsmath,mathrsfs}
\usepackage{hyperref}
\newcommand{\beq}{\begin{equation}}
\newcommand{\eeq}{\end{equation}}
\newcommand{\vp}{\vphantom}
\newcommand{\hp}{\hphantom}
\begin{document}
\begin{center}
{\Large\bf On integrability of $D0-$brane equations on $AdS_4\times\mathbb{CP}^3$ superbackground}\\[0.5cm]
{\large D.V.~Uvarov\footnote{E-mail: d\_uvarov@\,hotmail.com}}\\[0.2cm]
{\it NSC Kharkov Institute of Physics and Technology,}\\ {\it 61108 Kharkov, Ukraine}\\[0.5cm]
\end{center}
\begin{abstract}
Equations of motion for the $D0-$brane on $AdS_4\times\mathbb{CP}^3$ superbackground are shown to be classically integrable by extending the argument previously elaborated for the massless superparticle model.
\end{abstract}

\section{Introduction}

Integrable structures have become nowadays the basic objects of
study both in $AdS_5/CFT_4$ \cite{Maldacena} and
Aharony-Bergman-Jafferis-Maldacena (ABJM) \cite{ABJM} dualities.
In the former case classical integrability of the $AdS_5\times
S^5$ superstring equations was proved in the seminal work \cite{BPR}
based on the $AdS_5\times S^5$ superstring description as a $2d$
sigma-model on the $PSU(2,2|4)/(SO(1,4)\times SO(5))$ supercoset
manifold \cite{MT98}, \cite{Kallosh}. In the ABJM case theories
conjectured to be dual  share lower space-time (super)symmetry so
that integrable structure is more difficult to unveil and explore.
The gravity dual of ABJM gauge theory just in the special sublimit
of the 't Hooft limit reduces to the IIA superstring theory on
$AdS_4\times\mathbb{CP}^3$ superspace that is not isomorphic to a
supercoset manifold. Only its subspace of dimension $(10|24)$ can
be described as the $OSp(4|6)/(SO(1,3)\times U(3))$ supermanifold. The $OSp(4|6)/(SO(1,3)\times U(3))$ sigma-model \cite{AF}, \cite{Stefanski}\footnote{Alternative approach to
constructing the $OSp(4|6)/(SO(1,3)\times U(3))$ sigma-model
action relies on the introduction of the pure spinor variables
\cite{PS}.} corresponds to gauging away 8
fermionic coordinates for the supersymmetries broken by the
$AdS_4\times\mathbb{CP}^3$ superbackground in the complete
superstring action constructed in \cite{GSWnew}. 
In analogy with the $AdS_5\times S^5$ superstring case the $OSp(4|6)/(SO(1,3)\times U(3))$ sigma-model equations  
are classically integrable \cite{AF}, \cite{Stefanski} but 
integrability of the full set of $AdS_4\times\mathbb{CP}^3$
superstring equations which depend non-trivially on those 8
fermions is by no means obvious. However, in \cite{SW10} and
\cite{CSW} it was verified perturbatively in such fermionic
coordinates up to the second order that this indeed the case.

In the present and companion papers \cite{U-NPB}, \cite{U1212} we explore integrability of equations for the point-like dynamical objects in IIA superstring theory on $AdS_4\times\mathbb{CP}^3$ superbackground. In \cite{U1212} it was proved integrability of the equations of massless superparticle describing dynamics of the superstring zero modes. Here we extend the proof to the case of $D0-$brane, for which various formulations of the action and $\kappa-$symmetry gauge conditions were considered in \cite{GrSW}.

\section{Aspects of supergeometry of $AdS_4\times\mathbb{CP}^3$ superspace }

In this Section to make the presentation self-contained we sketch the construction of supervielbein and Ramond-Ramond (RR) 1-form on the $OSp(4|6)/(SO(1,3)\times U(3))$ supercoset manifold and the $AdS_4\times\mathbb{CP}^3$ superspace that are 'building blocks' of the $D0-$brane action with the emphasis on the relation to $D=3$ superconformal symmetry of the ABJM gauge theory \cite{BLS}. Thorough treatment of the supergeometry of $AdS_4\times\mathbb{CP}^3$ superspace is given in \cite{GSWnew}.

Using the isomorphism between $osp(4|6)$ superalgebra and $D=3$ $\mathcal N=6$ superconformal algebra left-invariant $osp(4|6)$ Cartan forms can be expanded over the $D=3$ $\mathcal N=6$ superconformal generators \cite{U08}
\beq
\label{osp46cf}
\begin{array}{rl}
\mathscr C(d)=\mathscr G^{-1}d\mathscr G=&\Delta(d)D+
\omega^m(d)P_m+c^m(d)K_m
+G^{mn}(d)M_{mn}\\[0.2cm]
+&\Omega_a(d)T^a+\Omega^a(d)T_a+\widetilde\Omega_a{}^b(d)\widetilde V_b{}^a+\widetilde\Omega_b{}^b(d)\widetilde V_a{}^a\\[0.2cm]
+&\omega^\mu_a(d)Q^a_\mu+\bar\omega^{\mu a}(d)\bar Q_{\mu a}+\chi_{\mu
a}(d)S^{\mu a}+\bar\chi^a_\mu(d)\bar S^\mu_a.
\end{array}
\eeq 
Bosonic subalgebra is spanned by $D=3$ conformal generators $(D, P_m, K_m, M_{mn})$ and $su(4)\sim so(6)$ $R-$symmetry generators that we divided into the $U(3)$ ones $\widetilde V_b{}^a$ forming the stability algebra of $\mathbb{CP}^3=SU(4)/U(3)$ manifold and the coset generators $(T_a, T^a)$. 24 fermionic generators correspond to $D=3$ $\mathcal N=6$ Poincare $(Q^a_\mu, \bar Q_{\mu a})$ and conformal $(S^{\mu a}, \bar S^\mu_a)$ supersymmetries. They are equipped with the $SL(2,\mathbb{R})$ spinor index $\mu=1,2$ and $SU(3)$ (anti)fundamental representation index $a=1,2,3$ in accordance with the decomposition $\mathbf 6=\mathbf 3\oplus\bar{\mathbf 3}$ of the $SO(6)$ vector on $SU(3)$ representations.

Geometric constituents of the $(10|24)-$dimensional
$OSp(4|6)/(SO(1,3)\times U(3))$ supermanifold that enter the $2d$
sigma-model equations and Lax connection \cite{AF},
\cite{Stefanski} are identified with the Cartan forms associated
with the generators $g_{(\mathrm k)}$, $\mathrm k=0,...,3$ having
definite eigenvalues w.r.t. the $\mathbb Z_4$ automorphism
$\Upsilon$ of the $osp(4|6)$ superalgebra: $\Upsilon(g_{(\mathrm
k)})=i^{\mathrm k}g_{(\mathrm k)}$, in particular invariant
eigenspace of $\Upsilon$ is spanned by the $so(1,3)\oplus u(3)$
stability algebra generators. So (\ref{osp46cf}) can be presented
in the manifestly $\mathbb Z_4-$graded form 
\beq\label{gradedcf}
\mathscr C(d)=\mathscr C_{(0)}(d)+\mathscr C_{(2)}(d)+\mathscr
C_{(1)}(d)+\mathscr C_{(3)}(d), 
\eeq 
where 
\beq
\begin{array}{rl}
\mathscr C_{(0)}(d)=&2G^{3m}(d)M_{3m}+G^{mn}(d)M_{mn}+\widetilde\Omega_a{}^b(d)\widetilde V_b{}^a+\widetilde\Omega_a{}^a(d)\widetilde V_b{}^b\in g_{(0)},\\[0.2cm]
\mathscr C_{(2)}(d)=&2G^{0'm}(d)M_{0'm}+\Delta(d)D+\Omega_a(d)T^a+\Omega^a(d)T_a\in g_{(2)},\\[0.2cm]
\mathscr C_{(1)}(d)=&\omega_{(1)}\vp{\omega}^\mu_{a}(d)Q_{(1)}\vp{Q}^a_\mu+\bar\omega^{\vp{\mu c}}_{(1)}\vp{\bar\omega}^{\mu a}(d)\bar Q_{(1)\mu a}\in g_{(1)},\\[0.2cm]
\mathscr C_{(3)}(d)=&\omega_{(3)}\vp{\omega}^\mu_{a}(d)Q_{(3)}\vp{Q}^a_\mu+\bar\omega^{\vp{\mu c}}_{(3)}\vp{\bar\omega}^{\mu a}(d)\bar Q_{(3)\mu a}\in g_{(3)}.
\end{array}
\eeq
The generators
\beq
M_{0'm}=\frac12(P_m+K_m),\quad M_{0'3}=-D,\quad M_{3m}=\frac12(K_m-P_m),\quad M_{mn}
\eeq
span the $so(2,3)$ algebra and one can write
\beq
g_{(0)}=(M_{3m},\ M_{mn}, \widetilde V_a{}^b),\quad g_{(2)}=(M_{0'm},\ D,\ T^a,\ T_a).
\eeq
Fermionic generators from $g_{(1)}$ and $g_{(3)}$ eigenspaces are defined by the linear combinations of those of Poincare and conformal supersymmetries
\beq\label{gradedfermidef}
Q^{\vp{a}}_{(1,3)}\vp{Q}^a_\mu=Q^a_\mu\pm iS^a_\mu,\quad\bar Q_{(1,3)}\vp{Q}_{\mu a}=\bar Q_{\mu a}\mp i\bar S_{\mu a}.
\eeq
As a result bosonic and fermionic Cartan forms from $\mathscr C_{(1,2,3)}$ eigenspaces
\beq\label{osp46bviel}
G^{0'm}(d)=\frac12(\omega^m(d)+c^m(d)),\quad\Delta(d),\quad\Omega_a(d),\quad\Omega^a(d)
\eeq
and
\beq
\omega_{(1,3)}\vp{\omega}^\mu_{a}(d)=\frac12(\omega^\mu_a(d)\pm i\chi^\mu_a(d)),\quad\bar\omega_{(1,3)}^{\hp{(1,3)}\mu a}(d)=\frac12(\bar\omega^{\mu a}(d)\mp i\bar\chi^{\mu a}(d))
\eeq
are identified with the $OSp(4|6)/(SO(1,3)\times U(3))$ supervielbein components, while remaining Cartan forms
\beq
G^{3m}(d)=\frac12(c^m(d)-\omega^m(d)),\quad G^{mn}(d),\quad\widetilde\Omega_a{}^b(d)
\eeq
describe the $SO(1,3)\times U(3)$ connection.

Since the $AdS_4\times\mathbb{CP}^3$ superspace cannot be realized as a supercoset manifold above described approach is not applicable directly to the construction of its geometric constituents that depend not only on the coordinates of the $OSp(4|6)/(SO(1,3)\times U(3))$ subsuperspace but also on 8 fermionic coordinates for the broken supersymmetries. However, the $AdS_4\times\mathbb{CP}^3$ background of IIA supergravity can be promoted to the maximally supersymmetric $AdS_4\times S^7$ background of $D=11$ supergravity thanks to the Hopf fibration realization of the 7-sphere $S^7=\mathbb{CP}^3\times S^1$ \cite{Nilsson}, \cite{STV}. The $AdS_4\times S^7$ superspace is isomorphic to the $OSp(4|8)/(SO(1,3)\times SO(7))$ supercoset manifold and one can construct $D=11$ supervielbein, connection and 4-form field strength out of the $osp(4|8)$ Cartan forms similarly to the above discussion albeit the $\mathbb Z_2$ automorphism of $so(2,3)\oplus so(8)$ subalgebra compatible with the cosetspace description of $AdS_4\times S^7$ space-time cannot be extended to the $\mathbb Z_4$ automorphism of the full $osp(4|8)$ superalgebra. Then performing dimensional reduction \cite{DHIS}, \cite{Howe} yields supervielbein and other constituents of the $AdS_4\times\mathbb{CP}^3$ supergeometry \cite{GSWnew}.

Using the isomorphism between $osp(4|8)$ superalgebra and $D=3$ $\mathcal N=8$ superconformal algebra left-invariant $osp(4|8)$ Cartan forms admit decomposition over the $D=3$ $\mathcal N=8$ superconformal generators \cite{U09}, \cite{U-IJMPA}
\beq\label{osp48cf}
\hat{\mathscr G}^{-1}d\hat{\mathscr G}=\hat{\mathscr C}_{\mathrm{so(2,3)}}+\hat{\mathscr C}_{\mathrm{so(8)}}+\hat{\mathscr C}_{\mathrm{32susy}},
\eeq
where
\beq \label{osp48-1}
\hat{\mathscr C}_{\mathrm{so(2,3)}}=\underline\Delta(d)D+\underline{\omega}^m(d)P_m+\underline c^m(d)K_m+\underline G^{mn}(d)M_{mn},
\eeq
\beq \label{osp48-2}
\begin{array}{rl}
\hat{\mathscr C}_{\mathrm{so(8)}}=&\Omega_a(d)T^a+\Omega^a(d)T_a+\widetilde\Omega_a{}^b(d)\widetilde V_b{}^a+\widetilde\Omega_b{}^b(d)\widetilde
V_a{}^a+h(d)H\\[0.2cm]
+&\widetilde\Omega_a(d)\widetilde T^a+\widetilde\Omega^a(d)\widetilde T_a+\Omega_a{}^4(d)V_4{}^a+\Omega_4{}^a(d)V_a{}^4
\end{array}
\eeq
and
\beq \label{osp48-3}
\begin{array}{rl}
\hat{\mathscr C}_{\mathrm{32susy}}=&\underline\omega^\mu_a(d)Q^a_\mu+\underline{\bar\omega}{}^{\mu a}(d)\bar Q_{\mu a}+\underline\chi{}_{\mu a}(d)S^{\mu a}+\underline{\bar\chi}{}^a_\mu(d)\bar S^\mu_a\\[0.2cm]
+&\omega^\mu_4(d)Q^4_\mu+\bar\omega^{\mu 4}(d)\bar Q_{\mu 4}+\chi_{\mu 4}(d)S^{\mu 4}+\bar\chi^4_\mu(d)\bar S^\mu_4.
\end{array}
\eeq
Eq.~(\ref{osp48-1}) contains $so(2,3)$ Cartan forms in conformal basis and Eq.~(\ref{osp48-2}) introduces Cartan forms for the $so(8)$ generators in a basis corresponding to the Hopf fibration realization of the 7-sphere $S^7=\mathbb{CP}^3\times S^1$. In such a basis one explicitly singles out generators of the $su(4)\oplus u(1)$ isometry algebra of $\mathbb{CP}^3\times S^1$. The generators $(\widetilde V_a{}^b, T_a, T^a)$ span the $su(4)$ algebra and commute with the $u(1)$ generator $H$. Remaining 12 generators $(\widetilde T_a, \widetilde T^a, V_a{}^4, V_4{}^a)$ belong to the $so(8)/(su(4)\times u(1))$ coset. Commutation relations and transformation to the conventional form of $so(8)$ generators are discussed in \cite{U09}, \cite{U-IJMPA} and rely on a particular convenient realization \cite{U10} for the K\" ahler 2-form on $\mathbb{CP}^3$ manifold.\footnote{For general consideration of the embedding of $su(4)$ algebra into $so(8)$ compatible with the Hopf fibration of the 7-sphere see \cite{GSWnew}.} The advantage of that choice of the K\" ahler 2-form consists in diagonalization of two projectors \cite{Nilsson}, \cite{GSWnew} used to divide 32 fermionic generators of the $osp(4|8)$ superalgebra (and associated coordinates) into 24 generators of the $osp(4|6)$ superalgebra and 8 generators corresponding to the supersymmetries broken by the $AdS_4\times\mathbb{CP}^3$ superbackground. Thus the first line in (\ref{osp48-3}) contains fermionic generators of $D=3$ $\mathcal N=6$ superconformal algebra while the second -- the generators of the broken supersymmetries.
%Altogether expressions (\ref{osp48-1})-(\ref{osp48-3}) describe embedding of the $osp(4|6)$ isometry superalgebra of $AdS_4\times\mathbb{CP}^3$
%%superbackground in the $osp(4|8)$ isometry superalgebra of $AdS_4\times S^7$ superspace.

$AdS_4\times S^7$ supervielbein components tangent to Anti-de Sitter part of the background equal to the $so(2,3)/so(1,3)$ Cartan forms 
\beq\label{11dvads}
\hat E^{m'}(d)=\left(\frac12(\underline{\omega}^m(d)+\underline c^m(d)),\ -\underline\Delta(d)\right),\quad m'=(m,3),
\eeq
and those tangent to $S^7$ in the basis corresponding to the Hopf fibration realization of the 7-sphere are given by
\beq \label{11dvs7}
E_{a}(d)=i(\Omega_a(d)+\widetilde\Omega_a(d)),\quad E^{a}(d)=i(\Omega^a(d)+\widetilde\Omega^a(d)),\quad\hat E^{11}(d)=h(d)+\widetilde\Omega_a{}^a(d),
\eeq
where 11-th space-time dimension is identified with the $S^1$ fiber.
Supervielbein fermionic components are identified with the odd Cartan forms in (\ref{osp48-3}). Explicit form of the supervielbein depends on the choice of $OSp(4|8)/(SO(1,3)\times SO(7))$ representative $\hat{\mathscr G}$. To fulfil the reduction to 10 dimensions supervielbein components should match the KK ansatz form \cite{DHIS}, \cite{Howe}. In the present case the reduction is performed along the $S^1$ parametrized by $y\in[0,2\pi)$ so that considering
\beq\label{scoset}
\hat{\mathscr G}=\mathscr Ge^{yH}\mathscr G_{\mathrm{br}}(\upsilon),
\eeq
where $\mathscr G\in OSp(4|6)/(SO(1,3)\times U(3))$ and $\mathscr G_{\mathrm{br}}$ is a function of 8 Grassmann coordinates for the broken supersymmetries $\upsilon_\mu=(\theta_\mu, \bar\theta_\mu, \eta_\mu, \bar\eta_\mu)$, ensures that the $osp(4|8)$ Cartan forms do not depend on $y$. Underlining in (\ref{osp48-1}), (\ref{osp48-3}) and (\ref{11dvads}) is used to indicate those of the $osp(4|6)$ Cartan forms that acquire dependence on $dy$, as well as  $\upsilon$ and $d\upsilon$ in addition to that on the coordinates of the $OSp(4|6)/(SO(1,3)\times U(3))$ supermanifold. In particular, the supervielbein components tangent to $AdS_4$ and $S^1$ acquire additive contributions proportional to $dy$
\beq\label{ads4s7viel'}
\hat E^{m'}(d)=G^{m'}(d)+G^{m'}_ydy,\quad\hat E^{11}(d)=\Phi dy+a(d),
\eeq
where $G_y^{m'}$ and $\Phi$ are functions of $\upsilon$. Expression (\ref{ads4s7viel'}) deviates from the Kaluza-Klein ansatz \cite{DHIS}, \cite{Howe} so that the $SO(1,4)$ Lorentz transformation $\mathsf L$ should be applied to remove the term proportional to $dy$ in $\hat E^{m'}(d)$
\beq
\mathsf L\left(
\begin{array}{c}
\hat E^{m'}(d)\\
\hat E^{11}(d)
\end{array}
\right)=\left(
\begin{array}{c}
E^{m'}(d)\\
\Phi_L(dy+A_L(d))
\end{array}
\right).
\eeq
Then $E^{m'}(d)$ is identified with the tangent to $AdS_4$ components of the $D=10$ supervielbein in KK frame,  $\Phi_L=\sqrt{\Phi^2+G_y^2}$ ($G^2_y=G_{ym'}G_y^{m'}\equiv G_y\cdot G_y$) is identified with $\exp{(2\phi/3)}$, where $\phi(\upsilon)$ is the $D=10$ dilaton
superfield, and $A_L$ is the Ramond-Ramond (RR) 1-form potential.
Explicit form of the entries of Lorentz rotation matrix is\footnote{Associated Lorentz rotation acting on the supervielbein fermionic components is discussed in \cite{GSWnew}.}
\beq\label{Lrot}
\mathsf L=\left(
\begin{array}{cc}
\delta^{m'}_{n'}+\frac{\Phi-\Phi_L}{\Phi_LG^{2\vp{\hat b}}_y}G^{m'}_yG_{yn'} & -\Phi_L^{-1}G^{m'}_y \\[0.2cm]
\Phi_L^{-1}G_{ym'} & \Phi_L^{-1}\Phi
\end{array}
\right)\in SO(1,4).
\eeq
$D=11$ supervielbein bosonic components $E_{a}(d)$ and $E^{a}(d)$ in (\ref{11dvs7}) do not depend on $dy$ and thus can be directly identified with the tangent to the $\mathbb{CP}^3$ components of $AdS_4\times\mathbb{CP}^3$ supervielbein in the KK frame, hence they were not endowed with 'hats' in (\ref{11dvs7}). This concludes characterization of the $AdS_4\times\mathbb{CP}^3$ supervielbein bosonic components and RR 1-form potential -- the constituents of the $D0-$brane action.

For the analysis of $D0-$brane equations in analogy with those of the massless superparticle it turns helpful to expand $G^{m'}(d)$, $E_{a}(d)$, $E^{a}(d)$ and $a(d)$ on the $\mathbb Z_4-$graded $osp(4|6)$ Cartan forms (\ref{gradedcf}) and $d\upsilon$ \cite{U1212}
\beq\label{expan1}
\begin{array}{rl}
G^{m'}(d)&=G^{0'n}(d)M_n{}^{m'}+G^{3n}(d)N_n{}^{m'}+\Delta(d)L^{m'}+G^{kl}(d)K_{kl}{}^{m'}\\[0.2cm]
+&q^{m'\mu}d\theta_\mu+\bar q^{m'\mu}d\bar\theta_\mu+s^{m'\mu}d\eta_\mu+\bar s^{m'\mu}d\bar\eta_\mu,\\[0.2cm]
E_a(d)&=i\Omega_a(d)+u_{(1)}\vp{u}^\mu\omega_{(1)\mu a}(d)+u_{(3)}\vp{u}^\mu\omega_{(3)\mu a}(d),\\[0.2cm]
E^a(d)&=i\Omega^a(d)+\bar u_{(1)}\vp{\bar u}^\mu\bar\omega_{(1)}\vp{\bar\omega}^a_\mu(d)+\bar u_{(3)}\vp{\bar u}^\mu\bar\omega_{(3)}\vp{\bar\omega}^a_\mu(d)
\end{array}
\eeq
and
\beq\label{expan2}
\begin{array}{rl}
a(d)=&\widetilde\Omega_a{}^a(d)+G^{0'm}(d)m_{m}+G^{3m}(d)n_{m}+\Delta(d)l+G^{mn}(d)k_{mn}\\[0.2cm]
+&h^\mu d\theta_\mu+\bar h^\mu d\bar\theta_\mu+p^\mu d\eta_\mu+\bar p^\mu d\bar\eta_\mu.
\end{array}
\eeq
Coefficients at the differentials of the 'broken' fermions $d\upsilon$ represent corresponding $AdS_4\times S^7$ supervielbein components while those at the $osp(4|6)$ Cartan forms can be named 'previelbeins'. Due to the choice of the supercoset element (\ref{scoset}) they are functions of $\upsilon$ only. To verify integrability of the $D0-$brane equations it is useful to have explicit expressions for them, as well as for $G_y^{m'}$ and $\Phi$ that can be derived by specifying $\mathscr G_{\mathrm{br}}$ (see \cite{U1212} for further details).

\section{$D0-$brane equations and their integrability}

The $D0-$brane action
\beq
S=-m\int d\tau\Phi_L^{-1}\sqrt{-\left(E_{\tau m'}E^{m'}_\tau-E_{\tau a}E_\tau\vp{E}^a\right)}+m\int A_L
\eeq
is the sum of kinetic  and Wess-Zumino (WZ) terms defined by the world-line pull-back of the RR 1-form potential $A_L$ with $m$ measuring its flux.
Introducing auxiliary $1d$ metric the action can be brought to the form without the square root
\beq\label{e-action}
S=\frac12\int d\tau\left[e^{-1}\Phi_L^{-2}\left(E_{\tau m'}E^{m'}_\tau-E_{\tau a}E_\tau\vp{E}^a\right)-em^2\right]+m\int A_L.
\eeq
Let us note at this point arbitrariness in the definition of the Lagrange multiplier $e(\tau)$. The choice made in (\ref{e-action}) gives conventional mass term $em^2$ independent of the superspace coordinates. It is also possible to 'absorb' $\Phi_L^{-2}$ factor into the definition of $e$ resulting in the action reducing in the $m\to0$ limit to that of the massless superparticle. The variation of (\ref{e-action}) on $e$ produces mass-shell constraint that is irrelevant for establishing integrability of other (dynamical) equations of motion and hence in the subsequent discussion we set $e=1/2$, i.e. concentrate on the '$1d$ sigma-model' case.

Considering the $osp(4|6)/(so(1,3)\times u(3))$ Cartan forms (\ref{gradedcf}) as independent variation parameters allows to derive the set of 10 bosonic and 24 fermionic equations of motion. They can be written as a system of the first order ordinary differential equations with the coefficients given by the world-line pullbacks of the $osp(4|6)$ Cartan forms to facilitate derivation of the Lax representation
\beq\label{beom}
\begin{array}{rl}
-\frac{\delta S}{\delta G^{0'}\vp{G}_m(\delta)}=&\dot a^{0'm}+2G_{\tau\hp{m}n}^{\hp{\tau}m}a^{0'n}-4l^{mn}G_{\tau\hp{0'}n}^{\hp{\tau}0'}+4fG_\tau\vp{G}^{3m}-2\Delta_\tau b^{3m}\\[0.2cm]
-&4i\left(\omega_{(1)\tau a}\sigma^m\bar\varepsilon_{(1)}{}^{a}-\varepsilon_{(1)a}\sigma^m\bar\omega_{(1)\tau}{}^a+\omega_{(3)\tau a}\sigma^m\bar\varepsilon_{(3)}{}^{a}-\varepsilon_{(3)a}\sigma^m\bar\omega_{(3)\tau}{}^a\right)=0,\\[0.2cm]
-\frac12\frac{\delta S}{\delta\Delta(\delta)}=&\dot f+G_{\tau\hp{0'}m}^{\hp{\tau}0'}b^{3m}-G_{\tau3m}a^{0'm}\\[0.2cm]
-&2\left(\omega_{(1)}\vp{\omega}^{\vp{\mu}}_\tau\vp{omega}^\mu_{a}\bar\varepsilon_{(1)}{}^{a}_\mu-\varepsilon_{(1)}{}^\mu_{a}\bar\omega_{(1)}\vp{\bar\omega}_\tau^{\vp{a}}\vp{\bar\omega}^a_\mu-\omega_{(3)}\vp{\omega}_\tau^{\vp{\mu}}\vp{omega}^\mu_a\bar\varepsilon_{(3)}{}^{a}_\mu+\varepsilon_{(3)}{}^\mu_{a}\bar\omega_{(3)}\vp{\bar\omega}_\tau^{\vp{a}}\vp{\bar\omega}^a_\mu\right)=0,\\[0.2cm]
-\frac{\delta S}{\delta\Omega_a(\delta)}=&\dot y^a+iy^b\left(\widetilde\Omega_{\tau b}\vp{\Omega}^a+\delta_b^a\widetilde\Omega_{\tau c}\vp{\Omega}^c\right)-4iw\Omega_\tau\vp{\Omega}^a\\[0.2cm]
+&4i\varepsilon^{abc}\left(\omega_{(1)}\vp{\omega}_\tau^{\vp{\mu}}\vp{omega}^\mu_b\varepsilon_{(1)\mu c}-\omega_{(3)}\vp{\omega}_\tau^{\vp{\mu}}\vp{omega}^\mu_b\varepsilon_{(3)\mu c}\right)=0
\end{array}
\eeq
and
\begin{equation}\label{feom}
\begin{array}{rl}
\frac14\frac{\delta S}{\delta\bar\omega_{(1)}\vp{\bar\omega}^a_\mu(\delta)}=&\dot\varepsilon_{(3)}{}^\mu_{a}\!+\!\frac12\!\left(G_\tau{}^{mn}\varepsilon_{(3)}{}^\nu_{a}\!-\! l^{mn}\omega_{(3)}\vp{\omega}_{\tau a}^{\hp{\tau}\nu}\right)\!\sigma_{mn\nu}{}^\mu\!+\! i\tilde\sigma_m^{\mu\nu}\!\left(G_\tau\vp{G}^{3m}\varepsilon_{(3)\nu a}\!-\!\frac{1}{2}b^{3m}\omega_{(3)}\vp{\omega}_{\tau\nu a}\right)\!\\[0.2cm]
+&i\tilde\sigma^{\mu\nu}_m\left(G_\tau\vp{G}^{0'm}\varepsilon_{(1)\nu a}-\frac{1}{2}a^{0'm}\omega_{(1)}\vp{\omega}_{\tau\nu a}\right)+\Delta_\tau\varepsilon_{(1)}{}^\mu_{a}- f\omega_{(1)}\vp{\omega}^{\hp{\tau}\mu}_{\tau a}\\[0.2cm]
-&i\!\left(\widetilde\Omega_{\tau a}\vp{\Omega}^b\!-\!\delta^b_a\widetilde\Omega_{\tau c}\vp{\Omega}^c\right)\!\varepsilon_{(3)}{}^\mu_{b}\!-\!2iw\omega_{(3)}\vp{\omega}^{\hp{\tau}\mu}_{\tau a}\!-\! i\varepsilon_{abc}\left(\Omega_\tau{}^b\bar\varepsilon_{(1)}{}^{\mu c}\!-\! y^b\bar\omega_{(1)}\vp{\bar\omega}_\tau^{\hp{\tau}\mu c}\right)=0,\\[0.2cm]
-\frac14\frac{\delta S}{\delta\bar\omega_{(3)}\vp{\bar\omega}^a_\mu(\delta)}=&\dot\varepsilon_{(1)}{}^\mu_{a}\!+\!\frac12\!\left(G_\tau{}^{mn}\varepsilon_{(1)}{}^\nu_{a}\!-\! l^{mn}\omega_{(1)}\vp{\omega}_{\tau a}^{\hp{\tau}\nu}\right)\!\sigma_{mn\nu}{}^\mu\!-\! i\tilde\sigma_m^{\mu\nu}\!\left(G_\tau{}^{3m}\varepsilon_{(1)\nu a}\!-\!\frac{1}{2}b^{3m}\omega_{(1)}\vp{\omega}_{\tau\nu a}\right)\!\\[0.2cm]
-&i\tilde\sigma^{\mu\nu}_m\left(G_\tau\vp{G}^{0'm}\varepsilon_{(3)\nu a}-\frac{1}{2}a^{0'm}\omega_{(3)}\vp{\omega}_{\tau\nu a}\right)+\Delta_\tau\varepsilon_{(3)}{}^\mu_{a}-f\omega_{(3)}\vp{\omega}^{\hp{\tau}\mu}_{\tau a}\\[0.2cm]
-&i\!\left(\widetilde\Omega_{\tau a}\vp{\Omega}^b-\delta^b_a\widetilde\Omega_{\tau c}\vp{\Omega}^c\right)\!\varepsilon_{(1)}{}^\mu_{b}\!-\!2iw\omega_{(1)}\vp{\omega}^{\hp{\tau}\mu}_{\tau a}\!-\! i\varepsilon_{abc}\left(\Omega_\tau{}^b\bar\varepsilon_{(3)}{}^{\mu c}\!-\! y^b\bar\omega_{(3)}\vp{\bar\omega}_{\tau}^{\hp{\tau}\mu c}\right)=0
\end{array}
\end{equation}
and c.c. equations. In Eqs.~(\ref{beom}), (\ref{feom}) the following quantities have been introduced
\beq\label{lso23def}
\begin{array}{rl}
\left(
\begin{array}{c}
 a^{0'}_{\hp{0'}m}\\
2f\\
-b_{3m}\\
-l_{kl}
\end{array}
\right)=&2\Phi_L^{-2}\left(
\begin{array}{c}
 M_m{}^{n'}\\
 L^{n'}\\
 N_m{}^{n'}\\
 K_{kl}\vp{K}^{n'}
\end{array}
\right)\{G_{\tau n'}-[\Phi^{-2}_L(G_\tau\cdot G_y+\Phi a_\tau)+\frac{m}{2}]G_{yn'}\}\\[0.3cm]
+&\Phi^{-2}_L\left(
\begin{array}{c}
 m_{m}\\
 l\\
 n_{m}\\
 k_{kl}
\end{array}
\right)[2\Phi_L^{-2}(G^2_ya_\tau-\Phi G_\tau\cdot G_y)+m\Phi],
\end{array}\eeq
\beq\label{lsu4def}
y^a=-i\Phi_L^{-2}E_\tau{}^a,\quad\bar y_a=-i\Phi_L^{-2}E_{\tau a},\quad w=\frac12\Phi^{-4}_L(G^2_ya_\tau-\Phi G_\tau\cdot G_y)+\frac14m\Phi_L^{-2}\Phi
\eeq
and
\beq\label{lfermidef}
\begin{array}{c}
\varepsilon_{(1)}{}^\mu_{a}=-\frac14\Phi_L^{-2}E_{\tau a}\bar u_{(3)}\vp{\bar u}^\mu,\quad\varepsilon_{(3)}{}^\mu_{a}=\frac14\Phi_L^{-2}E_{\tau a}\bar u_{(1)}\vp{\bar u}^\mu,\\[0.2cm]
\bar\varepsilon_{(1)}{}^{\mu a}=\frac14\Phi_L^{-2}E_\tau{}^au_{(3)}\vp{u}^\mu,\quad
\bar\varepsilon_{(3)}{}^{\mu a}=-\frac{1}{4}\Phi_L^{-2}E^{\vp{a}}_{\tau}\vp{E}^a_{\vp{tau}}u_{(1)}\vp{u}^\mu.
\end{array}
\eeq
In distinction to the superparticle case above introduced quantities take into account the WZ term contribution and the overall factor of $\Phi_L^{-2}$ in the kinetic term.  Finally there are 8 equations for the 'broken' fermions $\upsilon$ that can be brought to the following form
\beq\label{brfeom}
\begin{array}{rl}
-&2\frac{d}{d\tau}\left\{\Phi_L^{-2}\left[G_{\tau m'}+(\frac{m}{2}-\Phi_L^{-2}(G_\tau\cdot G_y+\Phi a_\tau))G_{ym'}\right]\frac{\partial G_\tau\vp{G}^{m'}}{\partial\dot\upsilon_\mu}+\Phi_L^{-2}[\frac{m}{2}\Phi\right.\\[0.2cm]
+&\left.\Phi_L^{-2}(G^2_ya_\tau-\Phi G_\tau\cdot G_y)]\frac{\partial a_\tau}{\partial\dot\upsilon_\mu}\right\}+\frac{\partial L}{\partial\Phi_L}\frac{\partial\Phi_L}{\partial\upsilon_\mu}
+\Phi_L^{-4}a_\tau\left[a_\tau\frac{\partial G^2_y}{\partial\upsilon_\mu}+(m\Phi_L^2-2G_\tau\!\cdot\! G_y)\frac{\partial\Phi}{\partial\upsilon_\mu}\right]\\[0.2cm]
+&\Phi_L^{-2}\left\{2G_{\tau m'}\frac{\partial G_\tau\vp{G}^{m'}}{\partial\upsilon_\mu}-E_{\tau a}\frac{\partial E_\tau^{\hp{\tau}a}}{\partial\upsilon_\mu}-E_\tau^{\hp{\tau}a}\frac{\partial E_{\tau a}}{\partial\upsilon_\mu}+[m-2\Phi_L^{-2}(G_\tau\!\cdot\! G_y+\Phi a_\tau)]\frac{\partial G_\tau\cdot G_y}{\partial\upsilon_\mu}\right. \\[0.2cm]
+&\left.[m\Phi+2\Phi_L^{-2}(G^2_ya_\tau-\Phi G_\tau\!\cdot\! G_y)]\frac{\partial a_\tau}{\partial\upsilon_\mu}\right\}=0,
\end{array}
\eeq
where the shorthand notation was introduced $\frac{\partial G_\tau\vp{G}^{m'}}{\partial\dot\upsilon_\mu}=(q^{m'\mu}, \bar q^{m'\mu}, s^{m'\mu}, \bar s^{m'\mu}),$\footnote{We assume that the fermionic derivative acts from the right.} $\frac{\partial a_\tau}{\partial\dot\upsilon_\mu}=(h^\mu, \bar h^\mu, p^\mu, \bar p^\mu)$ and
\beq
\begin{array}{rl}
\frac{\partial G_\tau\vp{G}^{m'}}{\partial\upsilon_\mu}=&G_\tau\vp{G}^{0'n}\frac{\partial M_n{}^{m'}}{\partial\upsilon_\mu}+G_\tau\vp{G}^{3n}\frac{\partial N_n{}^{m'}}{\partial\upsilon_\mu}+\Delta_\tau\frac{\partial L^{m'}}{\partial\upsilon_\mu}+G_\tau\vp{G}^{kl}\frac{\partial K_{kl}\vp{K}^{m'}}{\partial\upsilon_\mu}\\[0.2cm]
-&\dot\theta_\nu\frac{\partial q^{m'\nu}}{\partial\upsilon_\mu}-\dot{\bar\theta}_\nu\frac{\partial\bar q^{m'\nu}}{\partial\upsilon_\mu}
-\dot\eta_\nu\frac{\partial s^{m'\nu}}{\partial\upsilon_\mu}-\dot{\bar\eta}_\nu\frac{\partial\bar s^{m'\nu}}{\partial\upsilon_\mu},
\end{array}
\eeq
\beq
-\frac{\partial E_{\tau a}}{\partial\upsilon_\mu}=\omega_{(1)}\vp{\omega}_{\tau\nu a}\frac{\partial u_{(1)}\vp{u}^\nu}{\partial\upsilon_\mu}+\omega_{(3)}\vp{\omega}_{\tau\nu a}\frac{\partial u_{(3)}\vp{u}^\nu}{\partial\upsilon_\mu},\quad-\frac{\partial E_\tau^{\hp{\tau}a}}{\partial\upsilon_\mu}=\bar\omega_{(1)}\vp{\bar\omega}_{\tau\nu}^{\hp{\tau}a}\frac{\partial\bar u_{(1)}\vp{\bar u}^\nu}{\partial\upsilon_\mu}+\bar\omega_{(3)}\vp{\bar\omega}_{\tau\nu}^{\hp{\tau}a}\frac{\partial\bar u_{(3)}\vp{\bar u}^\nu}{\partial\upsilon_\mu}.
\eeq
Analogously are defined $\frac{\partial a_\tau}{\partial\upsilon_\mu}$ and $\frac{\partial G_\tau\cdot\ G_y}{\partial\upsilon_\mu}$.

Equations of motion (\ref{beom}), (\ref{feom}) and (\ref{brfeom}) are equivalent to the Lax equation
\beq\label{lax1}
\frac{d\mathscr L}{d\tau}+[\mathscr M,\mathscr L]=0
\eeq
with the Lax component $\mathscr M$ given by the world-line projection of the $osp(4|6)$ Cartan forms
(\ref{gradedcf}) and the component $\mathscr L$ can be presented as a sum
\beq\label{lax2}
\mathscr L=\mathscr L_{\mbox{\scriptsize so(2,3)}}+\mathscr L_{\mbox{\scriptsize su(4)}}+\mathscr L_{\mbox{\scriptsize24susys}}\in osp(4|6),
\eeq
where each summand is given by the linear combination of the quantities introduced in (\ref{lso23def})-(\ref{lfermidef})
\beq
\begin{array}{rl}
\mathscr L_{\mbox{\scriptsize so(2,3)}}=& a^{0'm}M_{0'm}+fD+b^{3m}M_{3m}+l^{mn}M_{mn},\\[0.2cm]
\mathscr L_{\mbox{\scriptsize su(4)}}=& y^aT_a+\bar y_{a}T^a+4w\widetilde V_a{}^a,\\[0.2cm]
\mathscr L_{\mbox{\scriptsize24susys}}=&\varepsilon_{(1)}{}^\mu_{a}Q^{\vp{a}}_{(1)}\vp{Q}^a_\mu+\bar\varepsilon_{(1)}{}^{\mu a}\bar Q_{(1)\mu a}+\varepsilon_{(3)}{}^\mu_{a}Q^{\vp{a}}_{(3)}\vp{Q}^a_\mu+\bar\varepsilon_{(3)}{}^{\mu a}\bar Q_{(3)\mu a}.
\end{array}
\eeq
$\mathscr L$ takes the same form as for the massless superparticle modulo the definition of coefficients (\ref{lso23def})-(\ref{lfermidef}).
Analogously to the superparticle case the Lax component $\mathscr L$ can be presented in the form of $osp(4|6)-$valued differential operator acting of the $D0-$brane action (\ref{e-action})
\beq\label{lax2compl}
\begin{array}{rl}
\mathscr L=&\!\!\left(M_{0'm}\frac{\partial}{\partial G_{\tau}\vp{G}^{0'}\vp{G}_m}+\frac12D\frac{\partial}{\partial\Delta_{\tau}}-M_{mn}\frac{\partial}{\partial G_{\tau mn}}-M_{3m}\frac{\partial}{\partial G_{\tau 3m}}+T_a\frac{\partial}{\partial\Omega_{\tau a}}+T^a\frac{\partial}{\partial\Omega_{\tau}\vp{\Omega}^a}\right.\\[0.2cm]
+&\!\!\left.\widetilde V_a{}^a\frac{\partial}{\partial\vp{\hat{\widetilde\Omega}}\widetilde\Omega_{\tau b}{}^b}-\frac14Q^{\vp{a}}_{(1)}\vp{Q}^a_\mu\frac{\partial}{\partial\bar\omega^{\vp{a}}_{(3)}\vp{\bar\omega}^{\hp{\tau}a}_{\tau\mu}}+\frac14\bar Q_{(1)\mu a}\frac{\partial}{\partial\omega_{(3)\tau\mu a}}+\frac14Q^{\vp{a}}_{(3)}\vp{Q}^a_\mu\frac{\partial}{\partial\bar\omega_{(1)}\vp{\bar\omega}^{\hp{\tau}a}_{\tau\mu}}-\frac14\bar Q_{(3)\mu a}\frac{\partial}{\partial\omega_{(1)\tau\mu a}}\right)\! S.
\end{array}
\eeq 

\section{Conclusion} 

In this note we centred in the proof of the classical integrability of equations of the 
$D0-$brane on $AdS_4\times\mathbb{CP}^3$
superbackground. They can be written as the system of first order
ordinary differential equations that admit Lax representation
with the Lax pair components taking value in the $osp(4|6)$
isometry superalgebra of $AdS_4\times\mathbb{CP}^3$
superbackground. This generalizes the proof of integrability of
the massless superparticle equations given in \cite{U1212}.
However, the most important problem is to find a generalization to
the superstring case, i.e. to find a zero curvature representation for
the $AdS_4\times\mathbb{CP}^3$ superstring equations. As follows
from the discussion in \cite{U-NPB} on the relation between the
superparticle's Lax pair and the Lax connection of the
superstring, the component $\mathscr L$ of the Lax pair captures the
form of the linear in $\ell_2$ contribution to the $2d$ Hodge
dualized part of the Lax connection up to insertions of the
functions $\ell_1$, $\ell_3$, $\ell_4$ of the spectral parameter
that is convenient to identify with $\ell_3$. It is thus interesting to
find out whether the form of the remaining part of the Lax connection
is also determined by the constituents of $AdS_4\times S^7$
supervielbein bosonic components and 'previelbeins' analogously to
$\mathscr L$.

\section{Acknowledgements}

The author is grateful to A.A.~Zheltukhin for stimulating discussions.

\end{document}